\documentclass[12pt,a4paper]{article}
\usepackage{epsfig}

\newcommand{\ie}{{\it i.e. }}
\newcommand{\beq}{\begin{eqnarray}}
\newcommand{\eeq}{\end{eqnarray}}

\begin{document}

\thispagestyle{empty}


\vskip 2cm

\begin{center}
{\Large\bf  Simple Pendulum Revisited}

\vskip 0.5cm

{\bf\it  Neha Aggarwal, Nitin Verma \& P. Arun\footnote{arunp92@yahoo.co.in,
arunp92@physics.du.ac.in}}
\vskip 0.25cm
{Department of Physics \& Electronics\\
S.G.T.B. Khalsa College, University of Delhi  \\
Delhi - 110 007\\
India} 
\vskip 0.5cm

\begin{abstract}
We describe a 8085 microprocessor interface developed to make reliable time
period measurements. The time period of each oscillation of a simple pendulum 
was measured using this interface. The variation of the time period with
increasing oscillation was studied for the simple harmonic motion (SHM) and
for large angle initial displacements (non-SHM). The results underlines the
importance of the precautions which the students are asked to take while 
performing the pendulum experiment. 

\end{abstract}	
\end{center}

\vskip 0.5cm

\section{Introduction}

\par The simple pendulum is a very trivial experiment that physics students do in 
higher secondary. Yet students sometimes fail to appreciate why "the initial 
angular displacement of the pendulum must be small". Interesting letter to the 
Editor in Physics Education (UK) point that the pendulum's time period 
increases by only 1\% for pendulum's oscillating through ${\rm 30^o}$. A 1\%
increase means the time period is 10msec more for than a pendulum undergoing
SHM whose time period is 1sec, thus students do question the very need of the 
precaution that the pendulum should only be given small angle initial 
displacements. Variations as small as 10msec are very difficult to measure
using stop-watches. Since computers (or microprocessors) can in principle
make measurements in micro-seconds, we were tempted to study the simple
pendulum using a micro-processor.

\section{The interface}

\par
\begin{figure}[htb]
\begin{center}
\epsfig{file=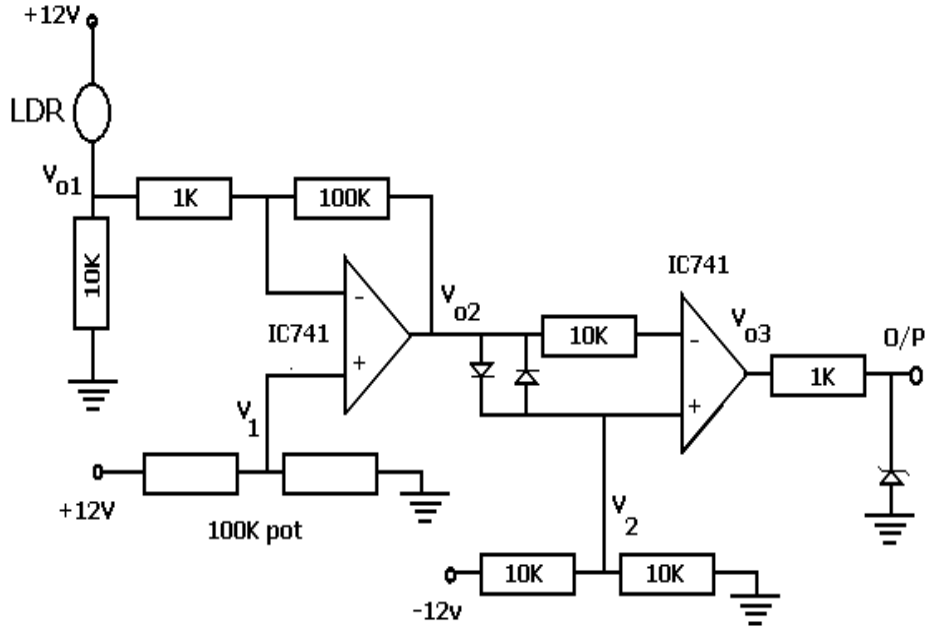,height=3.5in,width=5in}
\caption{Circuit used to wave-shape and interface instant of pendulum's
cross-over in front of the light dependent resistance (LDR).} 
\label{fig:1a}
\end{center}
\end{figure}

 The microprocessor is essentially made up of digital devices, which
communicate among itself in the language of binary, i.e. in ones (1) and
zeros (0). That is voltage signals of preferred levels. To find the time
interval of an oscillating pendulum, we keep an arrangement of laser source
and light dependent resistor (LDR) such that the pendulum's bob cuts the
light path during its oscillation. As the bob cuts the path, the light is
momentarily blocked. This produces a change in current generated in the LDR. 
For the microprocessor to communicate and understand this change (analog) in
current, it has to be converted to TTL compatible digital voltage. The
conversion and subsequent wave shaping is done using the circuitry shown in
fig(\ref{fig:1a}).

\par The value of resistance of the LDR, as the name suggests, depends on
whether light is falling on it or not. The resistance of the LDR is usually
inversely proportional to the intensity of light falling on it
\cite{street}. In our case when the laser light was falling on the LDR, it's
resistance was 15${\rm K\Omega}$, while on switching off the laser light it
rose to 150${\rm K\Omega}$. The voltage drop across the ${\rm 10K\Omega}$
resistance which forms a voltage divider with the LDR is 5v when the laser
light is 'ON' (bright phase) and 0.75v when LDR is not exposed to laser
(dark phase). This voltage (${\rm V_{o1}}$) could
have been directly fed into the microprocessor, however, the intensity
of the laser light strongly depends on the current supplied by it's
batteries. With time since the current is likely to fall, the voltage ${\rm
V_{o1}}$, would change as the experiment is being conducted. The difference 
amplifier, amplifies the difference between the voltages ${\rm V_{o1}}$ and 
${\rm V_1}$. By selecting a proper ${\rm V_1}$, using the pot, the output ${\rm
V_{o2}}$ varies from positive level for dark phase and negative level for
the bright phase. This inversion is bought about by the inverting amplifier
(opamp). The second opamp inverting action brings ${\rm V_{o3}}$ in phase
with ${\rm V_{o1}}$. This opamp is essentially a Schmitt's trigger, which
hard drives the output to +12v and -12v. The output of the second opamp 
(${\rm V_{o3}}$) varies from -12v to
+12v in accordance with the motion of the pendulum. Since, the
microprocessor can only understand zero or high state ($\sim$ +5v), a 4.7v 
zener diode is used to protect the the microprocessor (from +12v) by
converting +12v level to +4.7v and to force -12v level to zero. The ${\rm
1K\Omega}$ resistance kept between the opamp and zener diode is to control
the current flowing into the zener diode.

\begin{figure}[htb]
\begin{center}
\epsfig{file=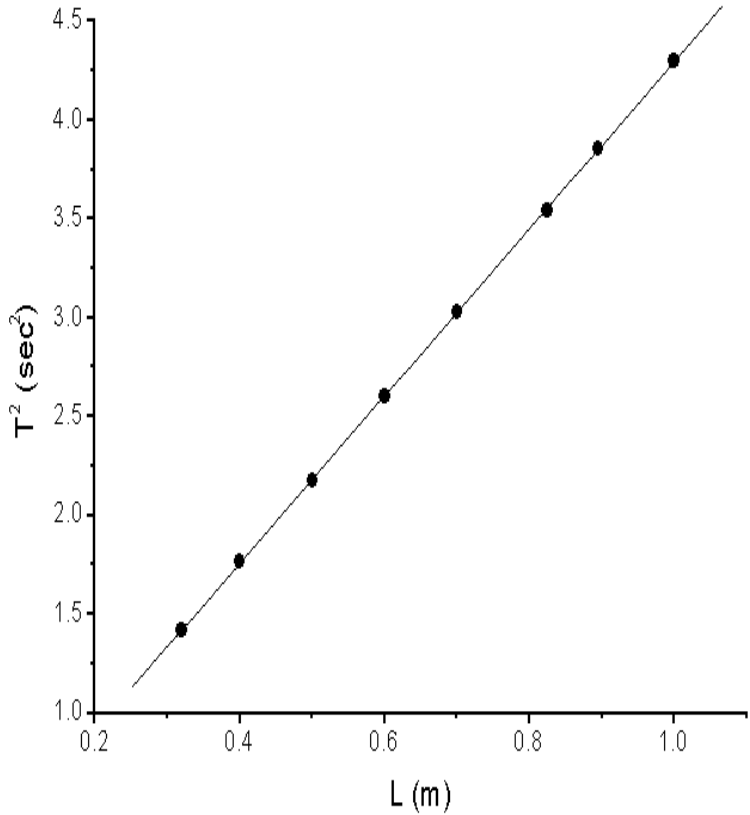,height=4.5in,width=4in}
\caption{Plot of pendulum length vs time period squared (${\rm
T^2}$). The data point fall on a straight line with co-relation factor as
good as 0.9995.} 
\label{fig:2}
\end{center}
\end{figure}
\par The realibilty of our microprocessor program (listed in the Appendix) 
was checked by finding the frequency of known square waves fed from audio 
function generators. The program is essentially a counter program which
counts the time interval taken between two positive edges of a train of
square waves. While counting, the program loops between instruction addresses
${\rm C024_H}$ and ${\rm C02D_H}$ for the square wave's high state and
between ${\rm C02E_H}$ and ${\rm C037_H}$ for the input square wave's ground
state. The latter part was to over come the requirement of IC555 monostable
trigger hardware in our circuitry. For a count of 'N', the time period is
given as 
\beq
T&=&{40N-3 \over f}\nonumber\\
&=&\left({40N-3 \over 3}\right)\mu s\label{form}
\eeq
where {\it 'f'} is the frequency of the microprocessor clock in MHz. The formula is
obtained by calculating the 'T' states or time taken for the microprocessor
to execute each instruction. It should be noted, the microprocessor's
program takes ${\rm 17\mu secs}$ to identify an edge and write the count in
a memory location. This ${\rm 17\mu secs}$ is a systematic error that would be 
present in the value of time period measured. From
this exercise we realized the quartz crystal used as a clock for the
microprocessor kit was 6.2MHz (Books \cite{goan} state it is 6MHz, however,
it depends on local manufacturers). Thus, results of all time measurements
(fig 2, 6, 8 and 11) have to be multiplied by 0.968(=6.0/6.2) to accommodate 
this correction. 

\par While the realibility of the software was established, to make
sure the LDR (our transducer), as also our microprocessor interface was
reliable, we measured the time period of oscillation for various pendulum 
lengths.
Figure (\ref{fig:2}) shows a perfect linearity between the pendulum's length
and it's time period squared (i.e. ${\rm T^2}$). This is expected and is
in accordance to well established theory that we shall discuss below. It
should be noted that on an average the response time of a LDR is in small
milli-secs. Thus, the ability to resolve and measure any changes in time
period with increasing oscillations would be in milli-secs. Also, the
accuracy in terms of absolute value of 'g' calculated from experimental data
would depend on how accurate the quartz crystals frequency has been
reported. 

\section{Simple Harmonic Motion: small initial displacement}

\par
\begin{figure}[htb]
\epsfig{file=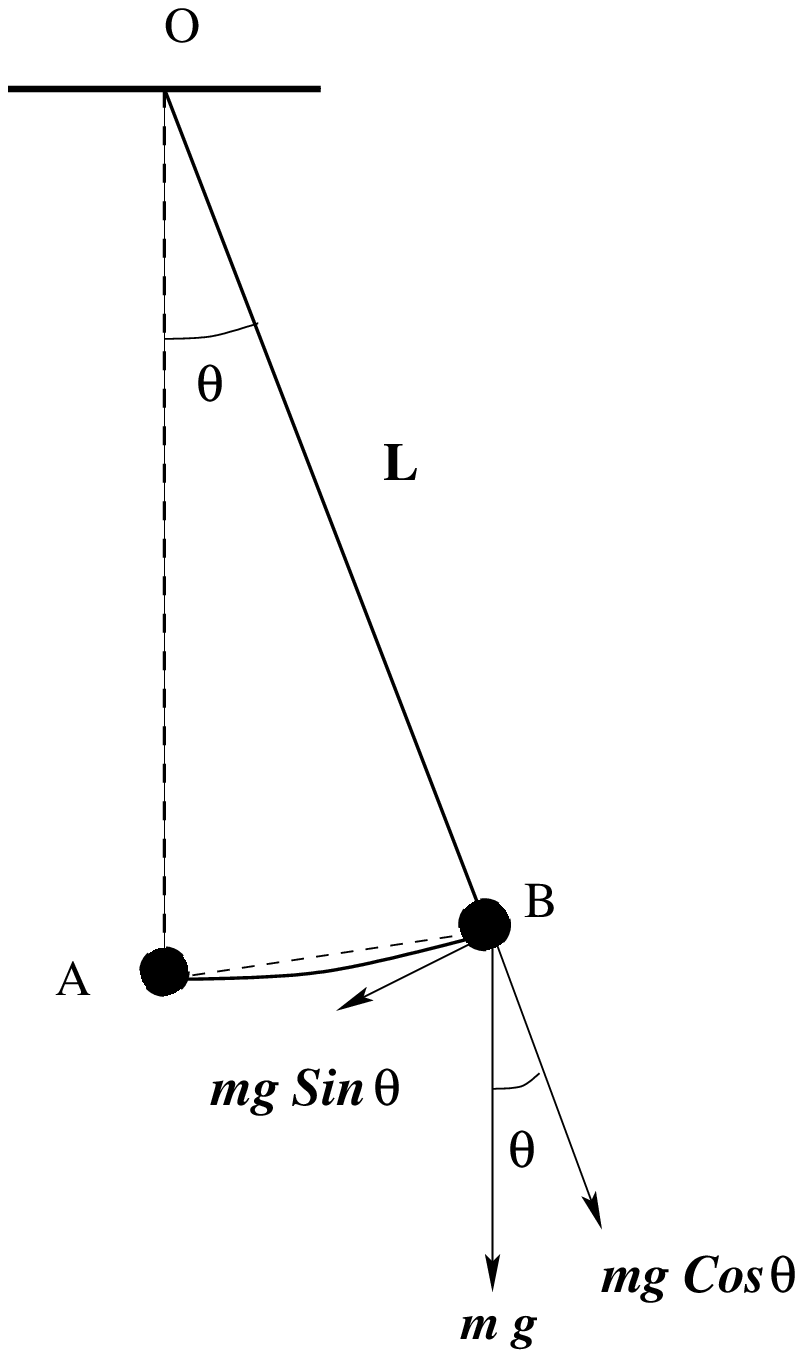,width=3in,height=4in} 
\hfil
\epsfig{file=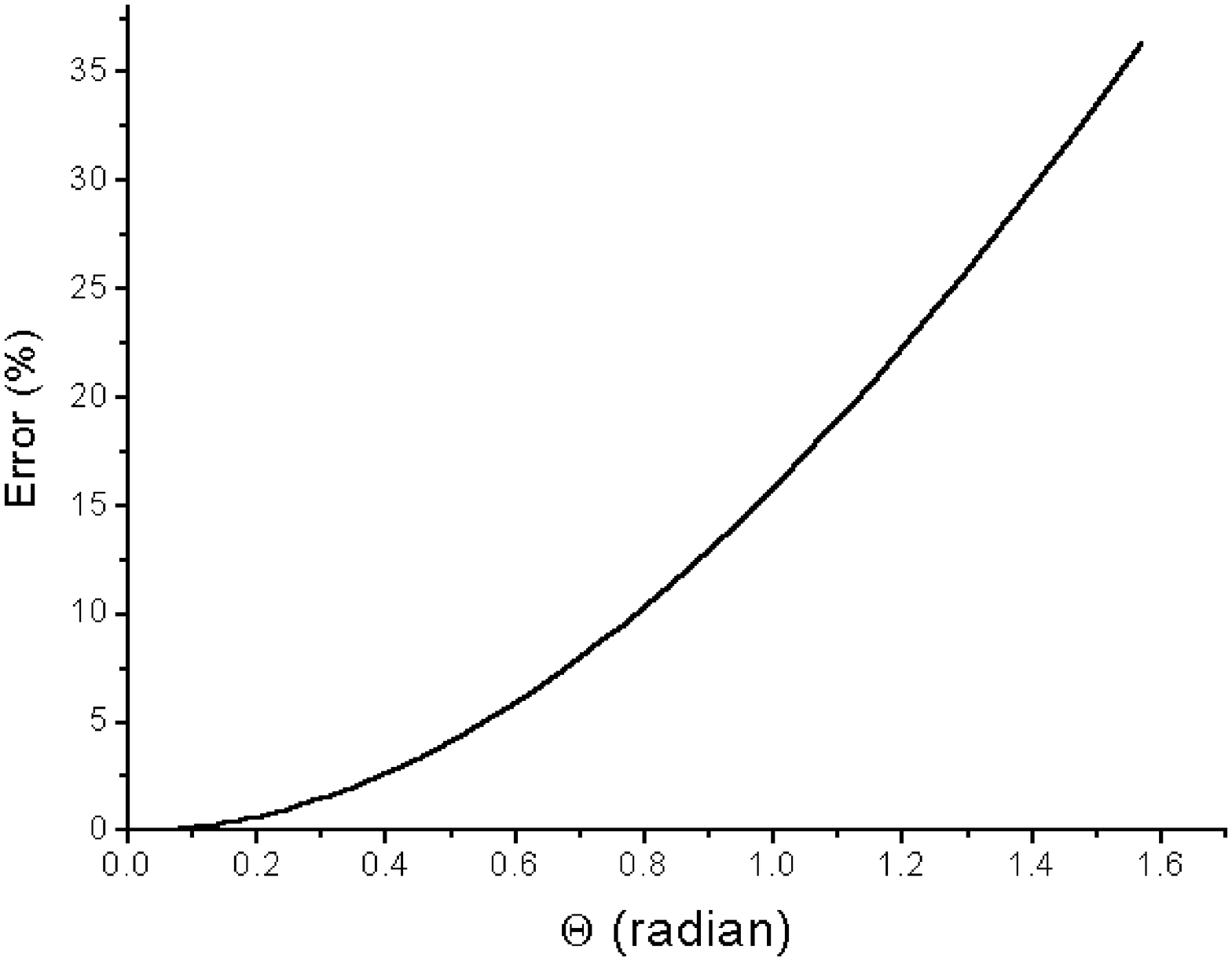,height=3in,width=4in}
\caption{Simple Pendulum of length L. Also shown is the plot of error (in \%) 
on assuming ${\rm sin \Theta = \Theta}$ increases with increasing $\Theta$.}
\label{fig:1}
\end{figure}

 Before proceeding to discuss the results of our experiment it would be
worthwhile to recapulate about the pendulum and under what conditions does
it's motion reduce to a simple harmonic motion. A pendulum is easily set up
by suspending a point mass. Physically, this is achieved by suspending a bob
which has an appreciable mass but whose radius is small as compared to the
length of the string used to suspend the bob in consideration. The pendulum
is set into to and fro motion by displacing it from it's mean position. The 
forces acting on the displaced pendulum is shown in fig(\ref{fig:1}). The 
restoring force is given as
\beq
F ~=~ -~ m g sin \Theta
\nonumber
\eeq
where {\it 'm'}, ${\rm '\Theta '}$ and {\it 'g'} are the mass of the bob, it's
angular displacement from the mean position, the acceleration due to
gravity respectively. The above leads to the equation of motion
\beq
m~L \frac{d^2\Theta}{dt^2} ~=~ -~m g sin \Theta
\nonumber
\eeq
or

\beq
\frac{d^2\Theta}{dt^2} ~~=~ - ~\frac{g}{L} ~sin\Theta
\label{eq:6}
\eeq
On considering the initial angular displacement \ie $\Theta$ to be small, 
${\rm sin\Theta}$ of eqn(\ref{eq:6}) reduces  to ${\rm \Theta}$ and 
substituting ${\rm \omega^2 ~=~ g/L}$, we have 
\beq
\frac{d^2\Theta}{dt^2} ~=~ - ~\omega^2 ~\Theta
\label{eq:7}
\eeq
This second order differential equation describes the motion of the simple 
harmonic motion (SHM), whose analytical solution is easily derivable and given 
as  
\beq
\Theta(t) ~=~ A~ sin(\omega t) ~+~ B~cos(\omega t)
\nonumber
\eeq
where A and B are constants. We can get the values of the constants by choosing
suitable initial conditions. The time period of oscillation can be obtained 
from the relationship ($ \omega ~=~ \frac{2 \pi}{T_o}$)   
\beq
\omega ~=~ \sqrt{\frac{g}{L}}
\nonumber
\eeq
giving
\beq
T_o=2\pi\sqrt{L \over g}
\label{eq:10}
\eeq

\par The above equation shows the proportionality between ${\rm T_o^2}$ and
the pendulum's length. To confirm the reliability of our time measuring
device (interface and software etc), we confirmed this relationship, see
fig(\ref{fig:2}). This relationship holds true for small angle
displacements. Hence, the data for fig(\ref{fig:2}) were collected for
various lengths of the pendulum with initial angular displacement being
${\rm 5^o}$, an universally accepted small angular displacement.

\par Students identify eqn(\ref{eq:10}) easily, since it is used by them to 
estimate 
the acceleration due to gravity. Also of interest is the fact that the above 
expression implies that the time taken to complete one oscillation is
independent of the angular displacement (${\rm \Theta}$), abide subjected to
the condition ${\rm sin \Theta \sim \Theta}$. It's here that the argument
starts as to what would be the appropriate initial displacement that a 
experimentalist should give to attain the simple harmonic motion? As ${\rm
\Theta }$ (in radians) increases, the disparity between itself and it's sine
(${\rm sin \Theta}$) increases. This fact is seen in fig(\ref{fig:1}), where
the increase in disparity is shown in terms of error (${\rm {\Theta -sin
\Theta \over \Theta}}$, in \%) w.r.t. ${\rm \Theta}$.

\par As can be seen, the error is below 10\% for angles less than ${\rm 45^o}$.
Would this limit be acceptable? Before answering this question, as to
understand the boundary between SHM and non-SHM, we proceed 
to understand the modifications introduced in eqn(\ref{eq:10}), when the
pendulum is set into motion with large angle displacements (non-SHM). 

\section{Pendulum with large initial displacement}

The time period of oscillation of a pendulum oscillating with large angles
can be found by solving eqn(\ref{eq:6}), i.e.
\beq
\frac{d^2\Theta}{dt^2} ~~=~ - ~\omega^2 ~sin\Theta
\label{non}
\eeq
However, that is easier said then done. Infact discussions on large
amplitude oscillations are rarely carried out because there are no
analytical solutions for the above differential equations. Infact, the
solution is expressed interms of elliptical integrals \cite{ma, smith}
\beq
T=\left({2 \over \pi}\right)T_o \int_0^{\pi/2}{d\Theta \over
\sqrt{1-sin^2(\Theta_m/2)sin^2\Theta}} 
\eeq
Hence, eqn(\ref{non}) is either numerically solved or various approximations 
are used. Of these approximations, the most famous was given by Bernoulli in
1749 \cite{smith}
\beq
T=T_o\left(1+{\Theta_m^2 \over 16}\right)
\label{eq:1}
\eeq
where ${\rm T_o}$ is the time period had the SHM condition been satisfied
and is given by eqn(\ref{eq:10}) and ${\rm \Theta_m}$ is the maximum angular
displacement given to the pendulum. Eqn(\ref{eq:1}) would suggest that take
whatever initial displacement you want while doing the experiment to
determine the acceleration due to gravity, all you have to do is to include
the correction (${\rm \Theta_m^2/16}$) in the time period expression (eqn
\ref{eq:6}).
\begin{figure}[htb]
\begin{center}
\epsfig{file=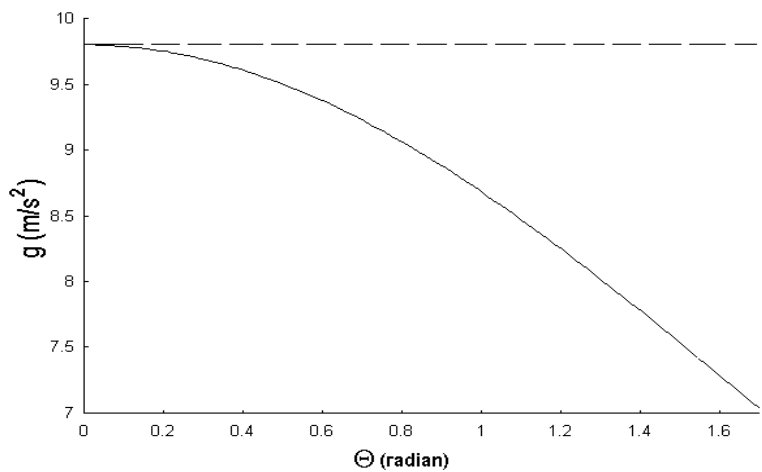,height=2.5in,width=4in}
\caption{Plot of acceleration due to gravity with initial displacement. The 
solid line is essentially calculated from 'T' of eqn(\ref{eq:1}) while the 
constant line is 'g' evaluated using ${\rm T_o={T \over 1+\Theta_m^2/16}}$ 
(i.e. time period after correcting for large angle displacement). }
\label{fig:7}
\end{center} 
\end{figure} 

\par While the pendulum and 
it's time period of oscillation in itself is interesting, it is usually used to 
evaluate the acceleration due to gravity. Teachers insist that students do the 
experiment with small angular displacements. Students do this without 
appreciating "why" and as to "what is a small angle". Let us consider what
kind of variation is expected theoretically if this precaution is not
adhered to. Figure(\ref{fig:7}) shows the plot 
of the variation of acceleration due to gravity with the initial displacement 
done by numerically solving eq(\ref{non}) \cite{pa}. The values of 'g' 
represented by the solid line is essentially calculated from 'T' of 
eqn(\ref{eq:1}) while the constant line is 'g' evaluated using ${\rm 
T_o={T \over 1+\Theta_m^2/16}}$ (i.e. time period after correcting for large 
angle displacement). As the figure shows, with increasing angular
displacement, the error in evaluated 'g' grows.

\par Eqn(\ref{eq:1}) suggests only a trivial consideration of including a 
correction factor is required if the small angle precaution is not followed, 
the question then arises "why fuss over small initial 
displacements?." Also, it is evident from fig(\ref{fig:7}) that the error in
'g' would be below 10\% for initial displacements below ${\rm 45^o}$ which
is quite a large angle. This might be well within the limits of experimental 
error, induced by your measuring devices like scale and stop-watch. It
should be noted that Nelson and Olsson \cite{nel} have determined 'g' with
an accuracy of ${\rm 10^{-4}}$ using a simple pendulum by including as many
as 8-9 correction terms. Thus, the importance 
of maintaining small initial displacement while performing the experiment 
is still not convincing. In the next section we report the measurements made
by our microprocessor interface and try to address the questions we have
asked above.

\section{Results \& Discussion}
\begin{figure}[htb]
\begin{center}
\epsfig{file=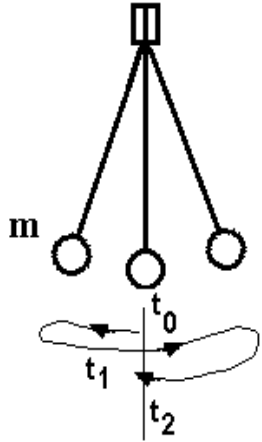,height=2.5in,width=1.5in}
\caption{Important instances for measuring one complete oscillation, with 
reference to our experiment.} 
\label{fig:pen2}
\end{center}
\end{figure}
\par A bob of radius 2.5cm was suspended using a cotton thread of length
one meter (the length of the pendulum thus is 1.025m). As the pendulum cut
the lasers path to the LDR, an electric pulse is generated. From the point
of the onset of the positive edge, the microprocessor counts the time elapsed
till the bob cuts the light path again. In one complete oscillation, the bob
cuts the light path thrice, say at instances ${\rm t_o}$, ${\rm t_1}$ and 
${\rm t_2}$ (see fig \ref{fig:pen2}). The time period is given as 
${\rm t_2-t_o}$. The program is
however designed to store ${\rm t_1-t_o}$ and ${\rm t_2-t_1}$. This was done
to make sure there is no error induced due to the inability to pin point the mean
position of the pendulum. Data were collected for the pendulum oscillating
with the initial displacements of ${\rm 5^o}$, ${\rm 10^o}$, ${\rm 15^o}$, ${\rm
20^o}$, ${\rm 25^o}$ and ${\rm 30^o}$. 
\begin{figure}[h]
\begin{center}
\epsfig{file=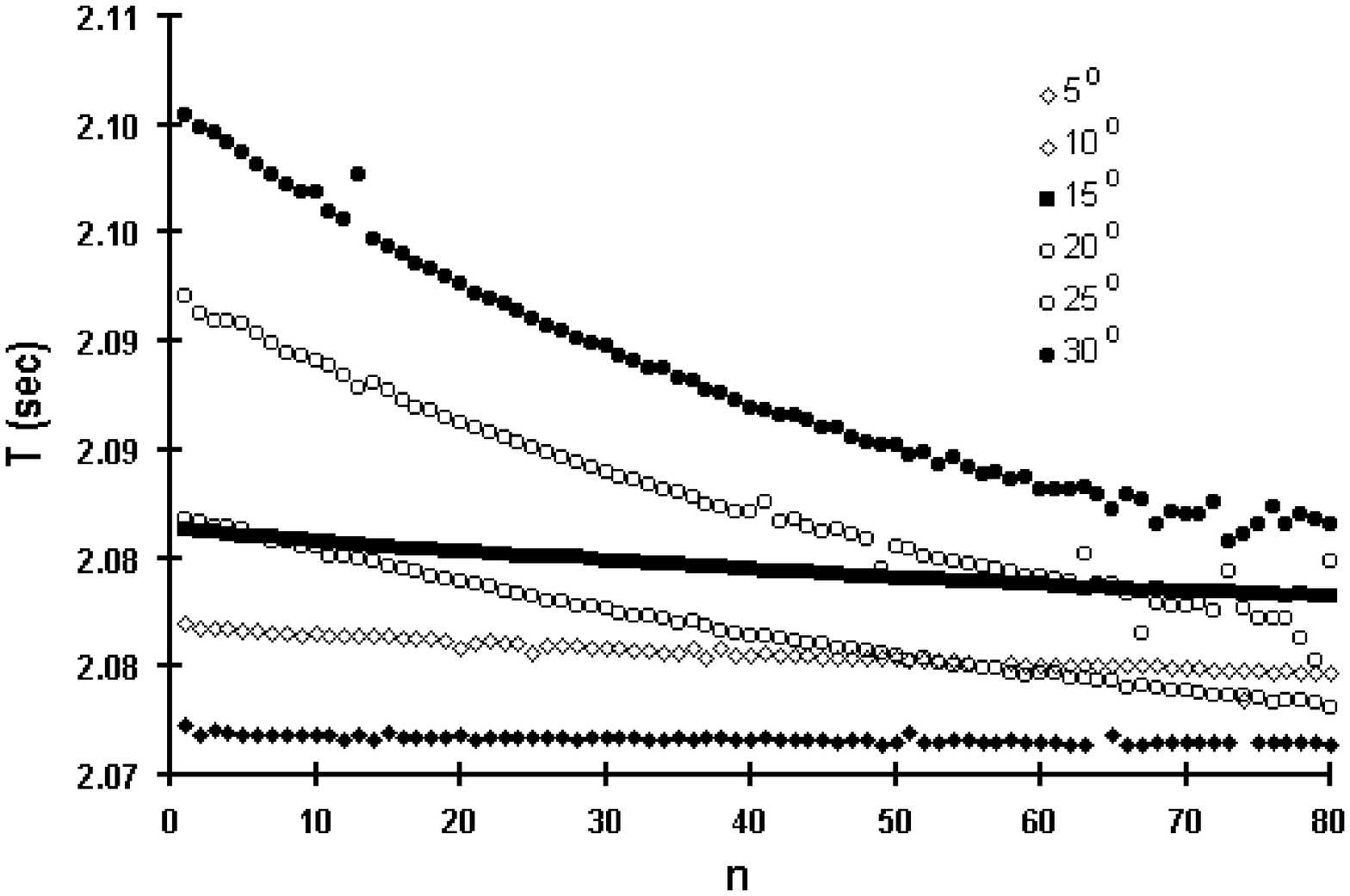,height=2.3in,width=3.2in}
\hfil
\epsfig{file=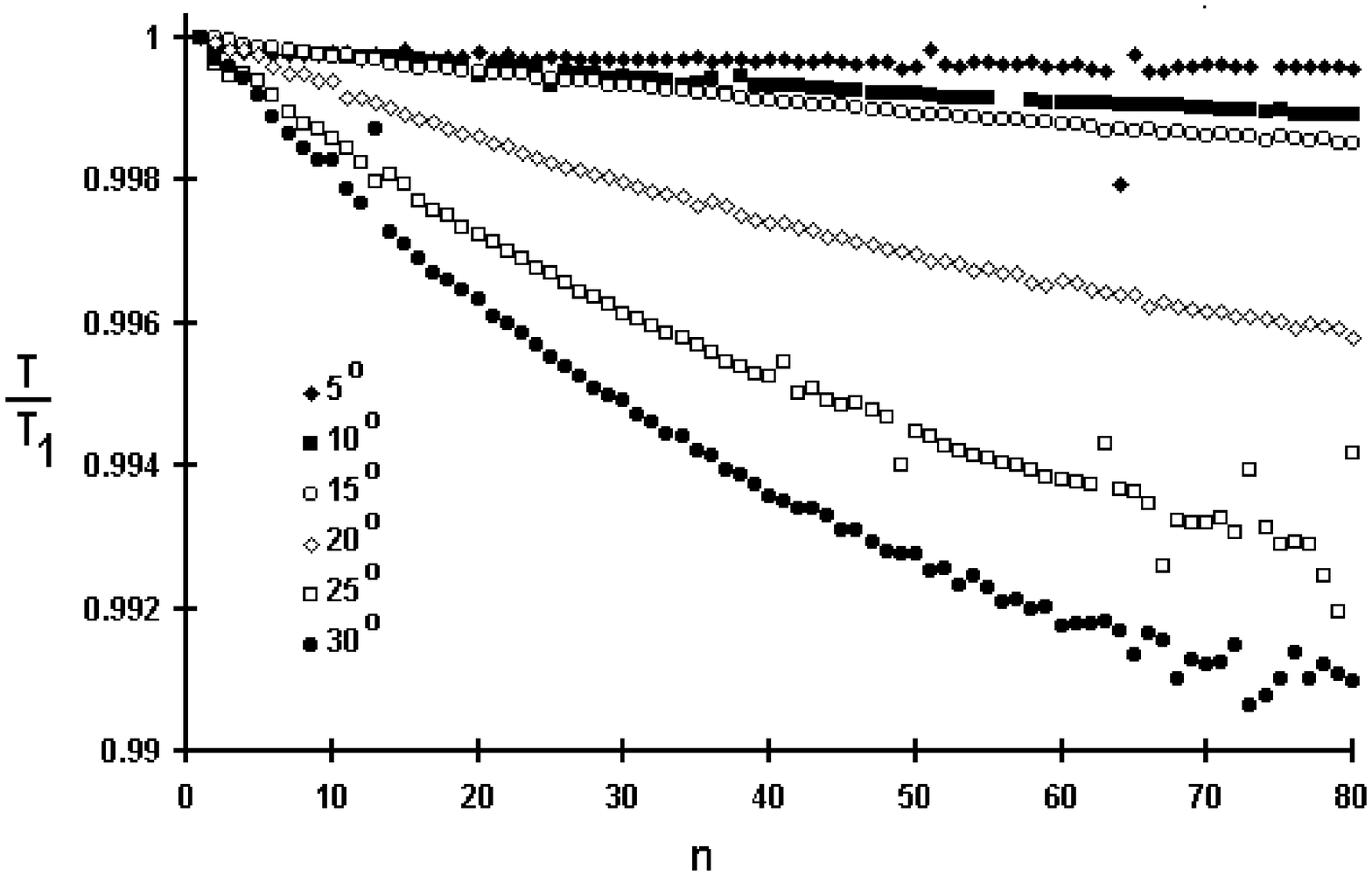,height=2.3in,width=3.2in}
\caption{Variation of the pendulum's time period with oscillations as
measured by the microprocessor interface. Along side is the normalized 
(${\rm T/T_1}$) time periods variation with oscillations.} 
\label{fig:ff1}
\end{center}
\end{figure}

\par Fig(\ref{fig:ff1}) shows the variation of the
pendulum's time period with oscillations. While for small initial angles,
there is no or slight variation in time period, for large initial
displacements, namely ${\rm 20^o}$, ${\rm 25^o}$ and ${\rm 30^o}$, the fall
in time period with increasing oscillations is pronounced. The fall is
better appreciated by plotting the normalized data (${\rm T/T_1}$). 
\begin{figure}[h]
\begin{center}
\epsfig{file=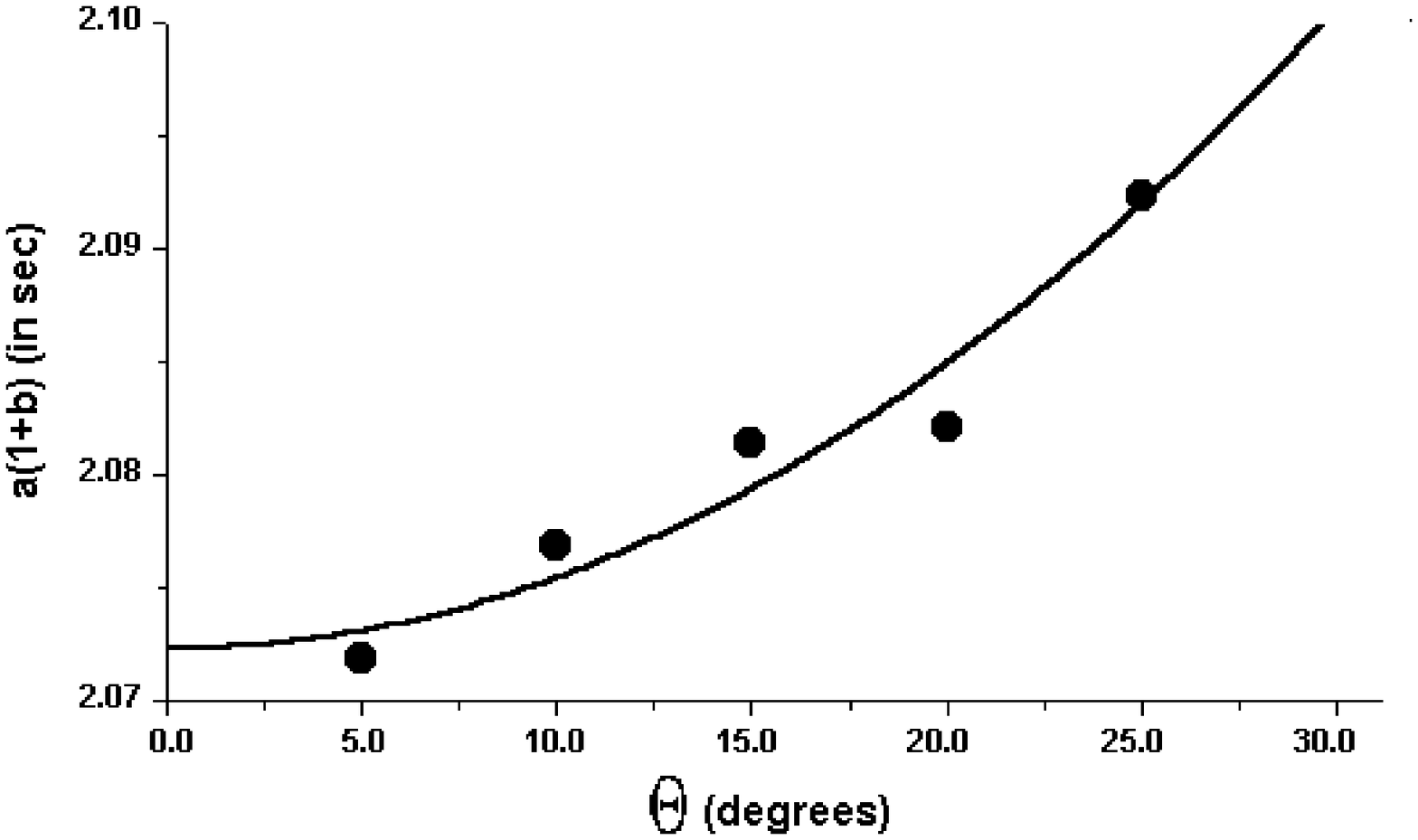,height=2.5in,width=3.5in}
\caption{Plot of ${\rm T(\beta =0)}$ (or a+ab) vs angle. The data points fit
well to Bernoulli's approx (eqn \ref{eq:1}) with co-relation factor as
good as 0.985.} 
\label{fig:a2}
\end{center}
\end{figure}
Deviation from the smooth variation in time period (scattering of data
points) is seen for large angles. This primarily is due to the pendulum's
support not being perfectly stationary. For this reason we restrict our
report to maximum angular displacement of ${\rm 30^o}$.

\par The fall in time period seems exponential. Since the experiment is not
ideal, one can expect damping to play a role (the decrease in amplitude was 
visible to the naked eye with increasing number of oscillations). Infact, 
damping is
expected to attenuate the amplitude of oscillation exponentially with time.
Thus, eqn(\ref{eq:1}) can be written as
\beq
T=T_o\left(1+{\Theta_m^2 e^{-2 \beta t} \over 16}\right)
\label{eq:d1}
\eeq  
Or, can be written as
\beq
T=a\left(1+b e^{-2 \beta t}\right)
\label{eq:d2}
\eeq  
Our objective would be to fit the above the equation to the
experimental data of fig(\ref{fig:ff1}). Table I lists the coefficient of
eqn(\ref{eq:d2}) obtained by curve fitting. 

\begin{center}
\vskip 0.2cm
{\bf Table I:} List of the coefficients obtained by fitting eqn(\ref{eq:d2})
to the experimental data of fig(\ref{fig:ff1}). The last column lists the
co-relation of fit with respective data points.
\vskip 0.2cm
\begin{tabular}{ccccccc}\hline\hline\\
S.No. & ${\rm \Theta}$ & a & b & ${\rm \beta}$ & ${\rm T(\beta =0)}$ & r \\
& (degrees) & (sec) & ${\rm (radian)^2}$ & ${\rm \times 10^{-2} sec^{-1}}$ &
(sec)\\ 
\hline\hline\\
1. & 5  & 2.00441  & 0.000333 & 0.34751 & 2.00508 & 0.8138 \\
2. & 10 & 2.00676  & 0.001552 & 0.31740 & 2.00988 & 0.9911 \\
3. & 15 & 2.01014  & 0.002072 & 0.38300 & 2.01430 & 0.9968 \\
4. & 20 & 2.00181  & 0.006540 & 0.32090 & 2.01490 & 0.9996 \\
5. & 25 & 2.00519  & 0.009755 & 0.41942 & 2.01038 & 0.9921 \\
6. & 30 & 2.00961  & 0.011860 & 0.50847 & 2.03344 & 0.9976 \\
\hline\hline\\
\end{tabular}
\end{center}


\par Instead of using
coefficient 'a' as a variable for curve fitting, it should be taken as a
constant (${\rm T_o}$, the SHM time period) defined by eqn(\ref{eq:10}).
That is, for all six values of ${\rm \Theta}$ (Table I), the value of 'a'
should work out to be the same. However, this proved to be a difficult
exercise where we were not able to achieve good co-relation between the
experimental data points and the fitted curve. To overcome this, after 
obtaining the generalized coefficients (a, b) we calculated ${\rm T(\beta =0)}$ 
(=a+ab, listed in Table I), i.e. the time period of the pendulum's
oscillation through large angles without any damping (the variation 
with angle of oscillation is 
given by Bernoulli's approximation eqn \ref{eq:1}). Fig(\ref{fig:a2}) shows
the variation of ${\rm T(\beta =0)}$ with angle. The solid line shows the
curve fit (eqn \ref{eq:1}). The ${\rm T_o}$ determined from our
experimental data works out to be 2.00545sec. One can use this value in
eqn(\ref{eq:10}) and determine the value of 'g', the acceleration due to
gravity. We get the acceleration due to gravity as ${\rm 9.963m/sec^2}$. 
This value is on the higher side. Even after accounting for the
influences corrections terms discussed in reference \cite{nel}, the value of
'g' would be on the higher side. The error, might be due to the lack of 
precise knowledge of the micro-processor's clock frequency. Using
eq(\ref{form}) and eq(\ref{eq:10}), we get the maximum possible error as
\beq
dg &=& \left({8 \pi^2 l \over T_o^2}\right){df \over f}\nonumber\\
&=& 6.54 \times 10^{-6} df\nonumber
\eeq
For an error of ${\rm +0.1 \times 10^6MHz}$ is the knowledge of the
micro-processor's clock frequency, the maximum error in 'g' would be
+0.65${\rm m/s^2}$. Thus the error, as stated, is due to the
imprecise knowledge of the micro-processor's clock frequency.

\begin{figure}[htb]
\begin{center}
\epsfig{file=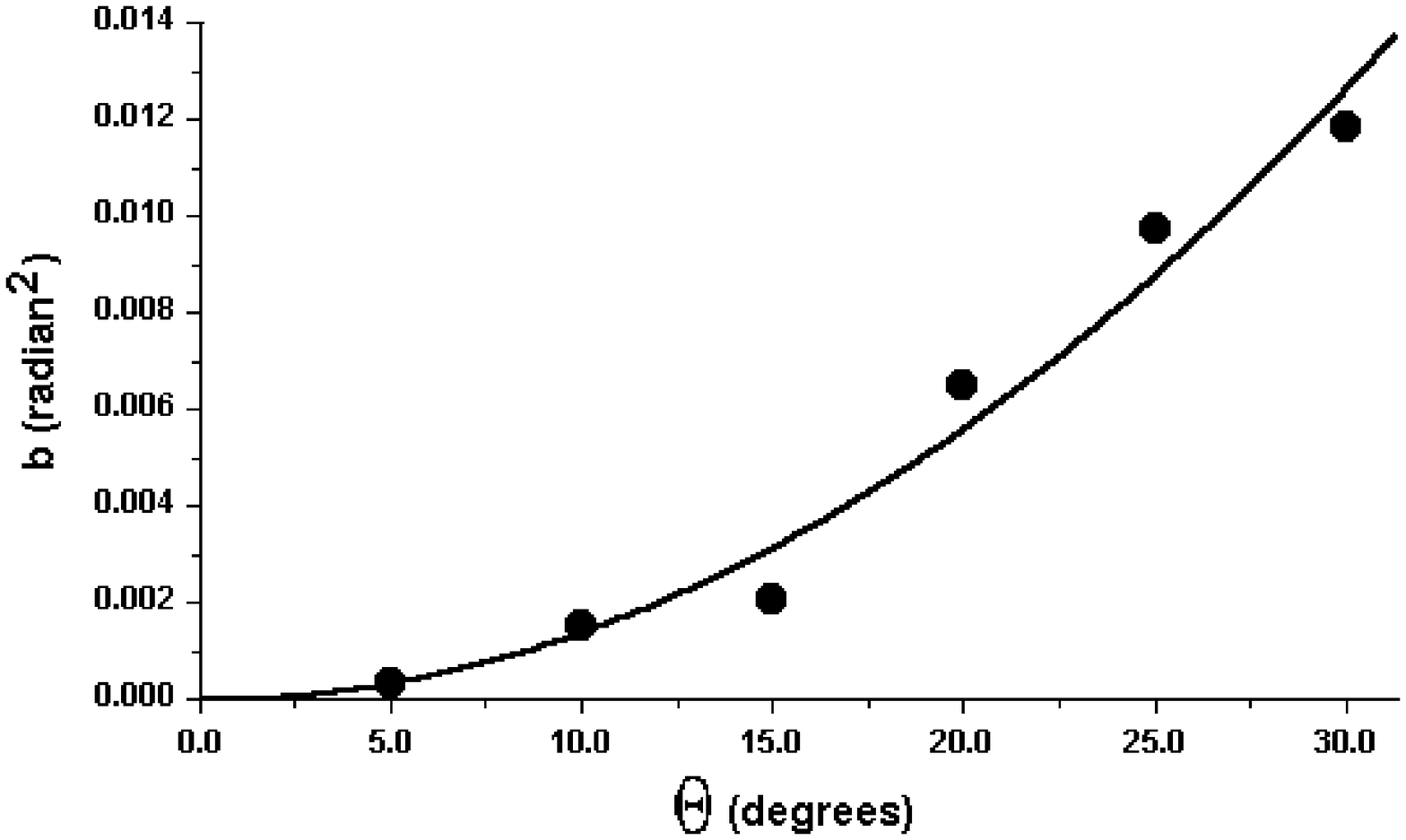,height=2.5in,width=3in}
\hfil
\epsfig{file=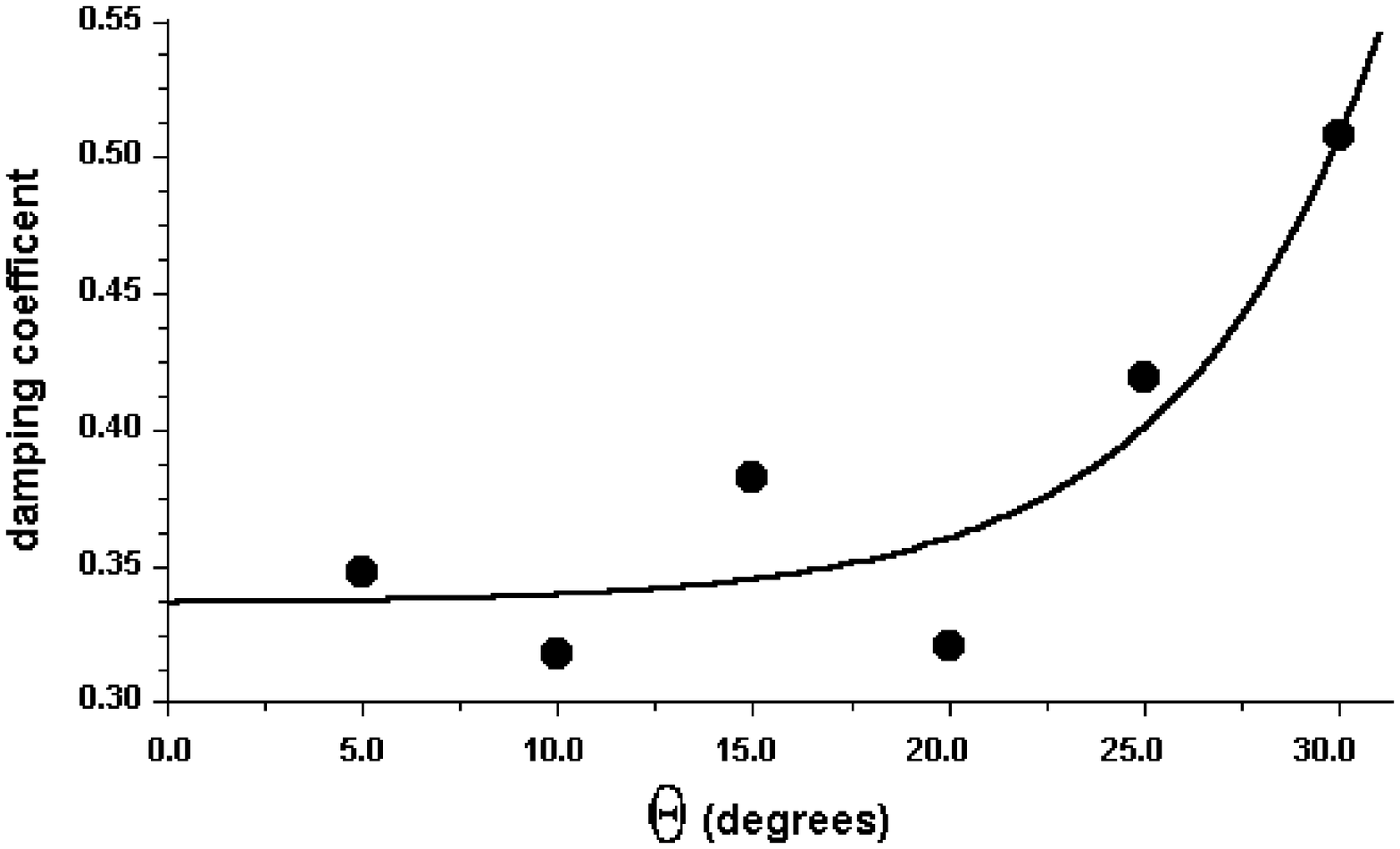,height=2.5in,width=3in}
\caption{Plot showing variation of coefficient 'b' with increasing initial
angular displacement. The solid line shows that the experimental data points
fall on a parabola. Second plot shows how the damping coefficient of the pendulum varies with
initial angular displacement.} 
\label{fig:d1}
\end{center}
\end{figure}
We now investigate the remaining coefficients 'b' and ${\rm \beta}$.
The coefficient 'b' is proportional to ${\rm \Theta^2}$ (compare
eqn \ref{eq:d1} and \ref{eq:d2}). This is evident from fig(\ref{fig:d1})
which shows the data points to fall nearly perfectly (co-relation factor is
0.984) on a parabola. The proportionality constant by eqn(\ref{eq:d1})
should be 1/16 (if ${\rm \Theta}$ is in degrees), or ${\rm 1.9 \times
10^{-5} rad^2}$. Our result gives the proportionality constant as 
${\rm 1.4 \times 10^{-5} rad^2}$ (or 1/21). Eqn(\ref{eq:d1}) is only an 
approximation hence, we can confidently
say that our data follows the solution given by Bernoulli (eqn \ref{eq:1}).

\begin{figure}[htb]
\begin{center}
\epsfig{file=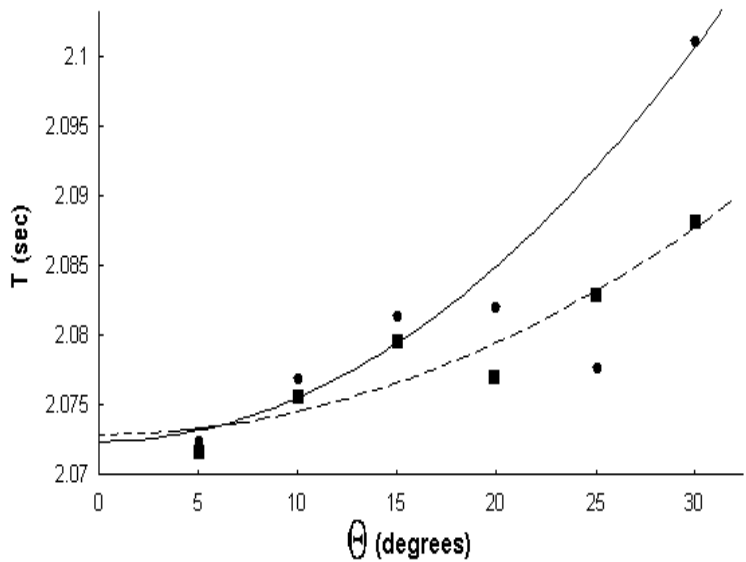,height=3in,width=4in}
\caption{Plot of ${\rm T(\beta=0)}$ and ${\rm T_{av}}$ vs angle. The solid
line is the curve fit of ${\rm T(\beta=0)}$ to eqn(\ref{eq:1}) while the dash
line is for visual aid to show variation of ${\rm T_{av}}$.} 
\label{fig:df1}
\end{center}
\end{figure}
\begin{figure}[htb]
\begin{center}
\epsfig{file=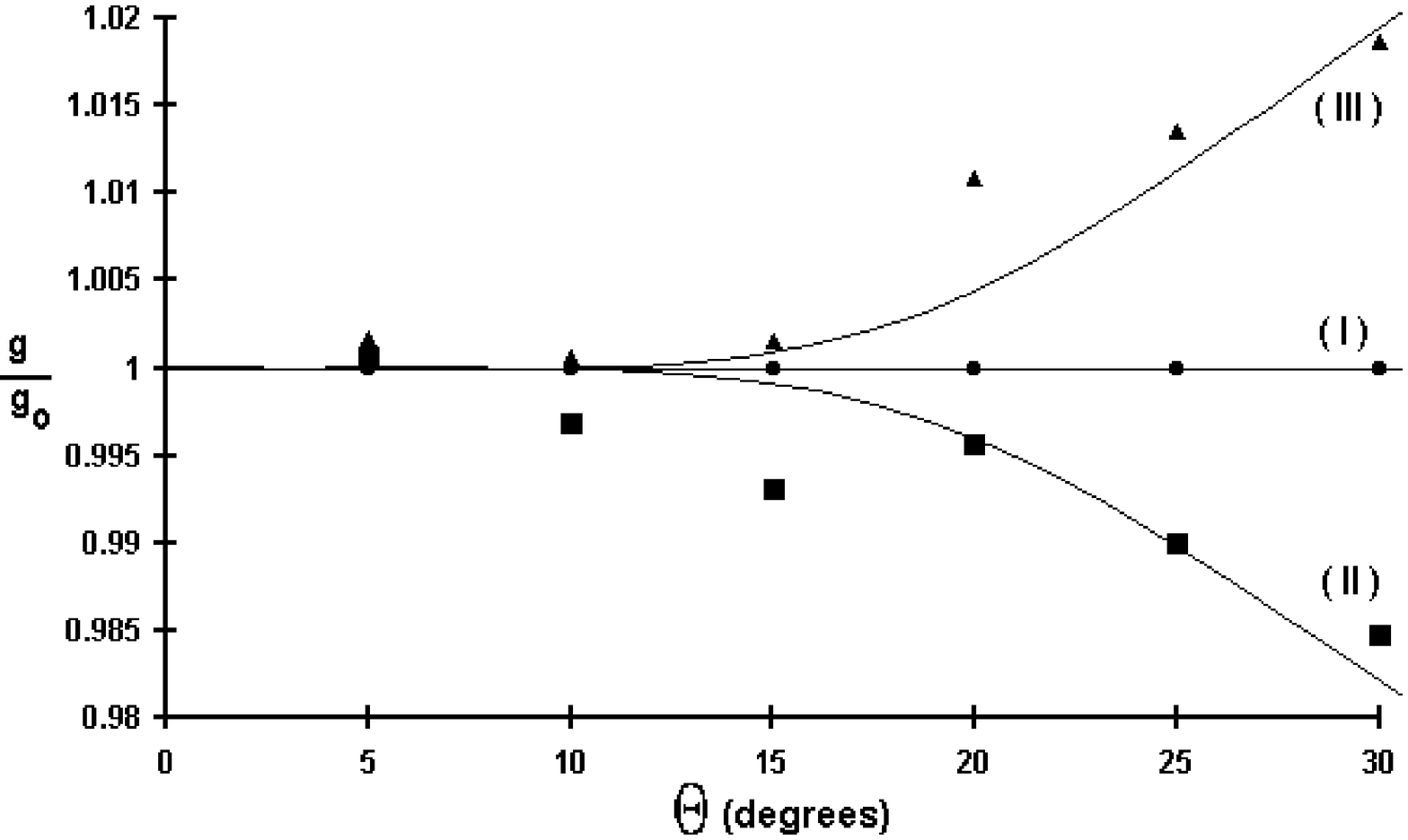,height=3in,width=4in}
\caption{Plot shows how the calculated acceleration due to gravity, 'g',
varies depending on what corrections have been done and it's variation with
initial angular displacement.} 
\label{fig:d2}
\end{center}
\end{figure}
\par The second graph of Fig(\ref{fig:d1}) shows the plot between the damping 
coefficient and the 
angular displacement. The damping factor is nearly constant for small angular
displacements which shows a rapid linear increase for angular displacements
above ${\rm 20^o}$. As can be seen the variation is similar to that of the IV
characteristics of a diode and as in it's case, we can extend the linear
region to cut the 'X' axis and look for the limiting initial angular displacement
which does not show sharp exponential fall in oscillation time period. This
works out to be ${\rm 11-12^o}$. Beyond this limit,
the damping coefficient is large and a pronounced exponential fall is seen
in the oscillation's time period (fig \ref{fig:ff1}). Kleppner and Kolenkow 
\cite{kk} have discussed the nature of ${\rm
\beta}$ and have stated that it depends on the shape of the mass and the
medium through which the mass moves. The amount of frictional force depends on 
the instantaneous velocity (${\rm d\Theta/dt}$) of the pendulum (${\rm
\beta}$ being the proportionality constant). However, this nature of the
frictional force (${\rm F=-\beta d\Theta/dt}$) is restricted for motion
where velocity is not large enough to cause turbulence. Beyond angles of
displacement of ${\rm 11-12^o}$, the frictional drag might not be following
the linear relationship with velocity. This however needs further
investigation.

\par Before summarizing the results of our experiment, it would be of use to
understand how the pendulum experiment is done in the undergraduate lab. The
student records the time taken to complete 40-50 oscillation \cite{cer}
oscillations from which the time period is calculated by dividing the total
time taken by the number of oscillations measured in that time. We call this
as ${\rm T_{av}}$ (average). This term obviously does not take into
consideration the influence of damping which is pronounced in large angle
oscillations. This obviously leads to erroneous results. Figure
(\ref{fig:df1}) compares ${\rm T_{av}}$ and ${\rm T(\beta=0)}$, the time
period after accounting for damping with the pendulum's displacement. The
figure clearly depicts the increasing disparity with large displacements.

\par Thus, if a student performs the pendulum experiment without taking the
necessary precaution of small angular displacement to get simple harmonic
motion and in turn ${\rm T_o}$, he or she would have to filter out the large
angle correction and the damping coefficient. If no correction is made and
'g' is calculated using ${\rm T_{av}}$ (listed in Table II), the variation
in 'g' with angle ${\rm \Theta}$, would increase (fig \ref{fig:d2} iii). The
fall in time period with successive oscillations is evident in this
experiment, since a micro-processor measures the time period. This would
not have been evident in ordinary circumstances. Thus, the experimenter
would not have been obvious of this and would only inco-operate large angle
corrections, with no corrections regarding damping. The resulting variation
is seen as the falling value of 'g' with angle in fig(\ref{fig:d2} ii). The
true constant nature of 'g' (fig \ref{fig:d2} ii) is only obtained when both
corrections are inco-operated.

\begin{center}
\vskip 0.2cm
{\bf Table II:} Listed is the  average time period ${\rm T_{av}}$ that a student 
would measure manually along with the time period if he bothers to correct
it for large angle oscillations. Also listed are the values of acceleration
due to gravity he would have got with his time periods.
\vskip 0.2cm
\begin{tabular}{cccccc}\hline\hline\\
S.No. & ${\rm \Theta}$ & ${\rm T_{av}}$ & ${\rm g_{av}}$ & ${\rm {T_{av}
\over 1+ \Theta^2/16}}$ & ${\rm g_{cor}}$  \\
& (degrees) & (sec) & (${\rm m/sec^2}$) & (sec) & (${\rm m/sec^2}$) \\ 
\hline\hline\\
1. & 5  & 2.00477  & 9.970 & 2.00382 & 9.979  \\
2. & 10 & 2.00866  & 9.931 & 2.00486 & 9.969  \\
3. & 15 & 2.01246  & 9.893 & 2.00389 & 9.978  \\
4. & 20 & 2.00980  & 9.920 & 1.99464 & 10.071  \\
5. & 25 & 2.01563  & 9.862 & 1.99197 & 10.098  \\
6. & 30 & 2.02093  & 9.811 & 1.98696 & 10.149  \\
\hline\hline\\
\end{tabular}
\end{center}

\section*{Conclusion}
By doing the pendulum experiment with large angle displacements,
calculations become complicated. As much as two informations have to be
filtered out, the effect of large angle displacement and the damping factor.
The damping coefficient is related to the initial displacement itself. These
informations can only be processed if the time period of each oscillation is
measured. This is quite impossible manually and only a micro-processor
interface is capable of highlighting these features.

\section*{Acknowledgement}
The authors would like to express their gratitude to the lab technicians of
the Department of Physics and Electronics, S.G.T.B. Khalsa College for the
help rendered in carrying out the experiments. 

\pagebreak
\subsection*{Appendix}
The microprocessor program required for measuring the time period of eighty
oscillations is listed below.
\begin{center}
\begin{tabular}{ccc}\hline\hline\\
Address & Instruction & Hex-code \\
\hline\hline\\
C000 & MVI E & 1E\\
C001 & ${\rm 160_D}$ & A0\\
C002 & LXI H & 21\\
C003 & 00 & 00\\
C004 & C1 & C1\\
C005 & MVI A & 3E \\
C006 & 00 & 00 \\
C007 & OUT & D3\\
C008 & 08 & 08 \\
C009 & IN & DB\\
C00A & 09 & 09\\
C00B & ANI & E6\\
C00C & 01 & 01 \\
C00D & CPI & FE\\
C00E & 00 & 00\\
C00F & JZ & CA \\
C010 & 15 & 15 \\
C011 & C0 & C0 \\
C012 & JMP & C3\\
C013 & 09 & 09\\
C014 & C0 & C0\\
C015 & IN & DB\\
C016 & 09 & 09\\
C017 & ANI & E6 \\
C018 & 01 & 01\\
C019 & CPI & FE\\
C01A & 01 & 01\\
C01B & JZ & CA \\
C01C & 21 & 21\\
C01D & C0 & C0 \\
C01E & JMP & C3\\
C01F & 15 & 15\\
\hline\hline\\
\end{tabular}
\end{center}
\vfill
\pagebreak
{\it continued ..}
\begin{center}
\begin{tabular}{ccc}\hline\hline\\
Address & Instruction & Hex-code \\
\hline\hline\\
C020 & C0 & C0\\
C021 & LXI B & 01\\
C022 & 00 & 00\\
C023 & 00 & 00\\
C024 & INX B & 03\\
C025 & IN & DB \\
C026 & 09 & 09\\
C027 & ANI & E6\\
C028 & 01 & 01\\
C029 & CPI & FE\\
C02A & 01 & 01\\
C02B & JZ & CA\\
C02C & 24 & 24\\
C02D & C0 & C0\\
C02E & INX B & 03\\
C02F & IN & DB\\
C030 & 09 & 09\\
C031 & ANI & E6\\
C032 & 01 & 01\\
C033 & CPI & FE\\
C034 & 00 & 00\\
C035 & JZ & CA\\
C036 & 2E & 2E\\
C037 & C0 & C0\\
C038 & MOV M, B & 70\\
C039 & INX H & 23\\
C03A & MOV M, C & 71\\
C03B & INX H & 23\\
C03C & DCR E & 1D\\
C03D & MOV A, E & 7B\\
C03E & CPI & FE\\
C03F & 00 & 00\\
C040 & JNZ & C2\\
C041 & 21 & 21\\
C042 & C0 & C0\\
C043 & HLT & 76\\
\hline\hline\\
\end{tabular}
\end{center}

\vfill
\pagebreak

\vfill

\end{document}